\documentstyle[11pt,IAU207_pasp,twoside,psfig]{article}
\begin{document}
\title{Dwarf Galaxies and Star Clusters in Tidal Tails}
 \author{Peter~M.~Weilbacher, Uta~Fritze-von~Alvensleben}
\affil{Universit\"ats-Sternwarte G\"ottingen, Geismarlandstr.~11, 37083 G\"ottingen, Germany}
\author{Pierre-Alain~Duc}
\affil{CNRS and CEA/DSM/DAPNIA/SAp, Saclay, 91191 Gif sur Yvette cedex, France}

\begin{abstract}
  We summarize the properties of tidal dwarf candidates in a sample of
  interacting galaxies and classify objects in tidal tails depending
  on their morphological appearance. New high-resolution dynamical
  models are needed to understand how the different structures seen in
  tidal tails are formed.
\end{abstract}

\section{Introduction}
First we want to stress the new definition for Tidal Dwarfs given by
Weilbacher \& Duc (2001), who -- following Duc et al.~2000 -- define a
Tidal Dwarf Galaxy (TDG) as a self-gravitating entity of dwarf-galaxy
mass built from tidal material expelled during interactions. To prove
that a given `knot' in or near a tidal feature is a real TDG, one
therefore has to show several properties:
\begin{itemize}
\item Luminosity (and therefore mass) of a dwarf galaxy.
\item Association with the interacting system.
\item Internal kinematics decoupled from the environment.
\item Not a preexisting dwarf galaxy involved in the interaction.
\item The object has to have a possible future existence to call it an `entity'.
\end{itemize}
The luminosity and the scale length of objects that this definition
allows therefore clearly separate TDGs from young or old star
clusters, which do not show rotation and have lower luminosities and
masses.

Typical interacting systems where gas-rich, star-forming TDGs are
believed to have been found are NGC 7252 (Hibbard \& Mihos 1995), Arp
242 (Duc et al., in prep), Arp 105 (Duc et al.~1997), and Arp 245 (Duc
et al.~2000). A formation sequence outlining a possible way these
objects are formed, has been given by Weilbacher \& Duc (2001) based
on existing numerical simulations: The condensations can first appear
in the gaseous or stellar component, and the other component is then
pulled into the potential well.  Collapsing gas could then initiate
the observed star formation in these tidal dwarfs.

\section{Summary of TDG properties}
We have analyzed a sample of 14 interacting systems (10 of them are
discussed in Weilbacher et al.~2000). About 40 knots in the tidal
features of these systems show properties, which make them plausible
TDG candidates.  They are brighter than $-11\,$mag in $M_B$, mostly
low surface brightness, and are extended. This differentiates them
from possible clusters and meets the first property listed above.
They also show blue colors, and are sometimes the bluest part of the
entire interacting system, which hints to active star formation.
Indeed they also show emission line spectra, which enables us to
measure the redshift and the metallicity in the gas phase. Several of
these TDG candidates also show signs of dynamics on their 2D spectra,
indicative of internal kinematics.

From comparison with evolutionary synthesis models (see Weilbacher \&
Fritze-v.Alvensleben 2001) we were able to estimate burst ages and
strengths. It was seen from the broad-band colors that these TDG
candidates indeed experience strong bursts increasing the stellar mass
by up to 20\%, and have young burst ages up to 20 Myrs (Weilbacher et
al.~2000). Preliminary mass estimates based on optical luminosities
show stellar masses of several $10^7$ M$_\odot$ or more.

An indication of `normal` star formation in TDGs has been found by
Lisenfeld et al.~(2001) by the detection of CO in 8 of 11 TDGs of
interacting systems in a different sample, with an efficiency similar
to that in spiral galaxies.

\section{Structures in Tidal Tails}
Looking at images of tidal features one can see different structures
in the tails (see also Hibbard 2001):
\begin{figure}
  \centerline{\vbox{
      \psfig{figure=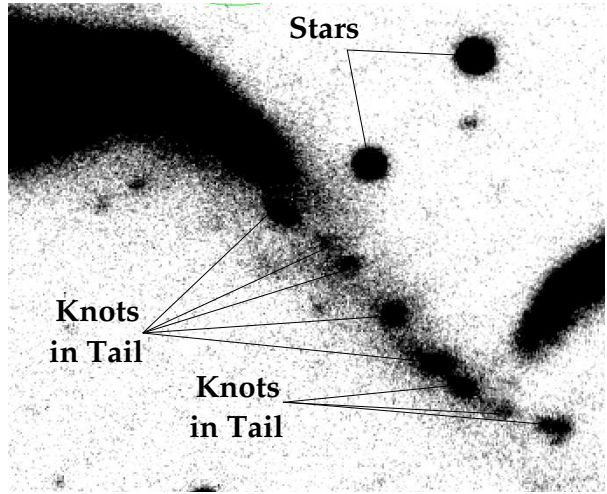,width=0.6\linewidth,angle=0}
  }}
  \caption{$B$-band image of the southern tail of AM 1353-272.}
\end{figure}
\begin{enumerate}
\item {\bf TDG(s)}: One or two tidal dwarfs per systems seem to be
  located preferably at the end of tidal tails. Not much structure
  elsewhere in the same tail. Examples: NGC 7252W (Knierman et
  al.~2001), AM 1159-530 (Weilbacher \& Duc 2001).
\item {\bf SSCs/GCs}: Lots of star clusters are seen along the tidal
  tail, age and association not well restricted. Example: Western tail
  of NGC 3256 (Knierman et al.~2001).
\item {\bf Extended knots}: A chain of extended knots are seen along
  the tidal tail. They are not resolved into HII regions, and
  therefore represent an intermediate case.  Example: Both tails of AM
  1353-272 (Fig.~1).
\end{enumerate}
The creation of TDGs (case~1) has been successfully modeled
dynamically, although the difference between stellar (Barnes \&
Hernquist 1992) and gaseous (Elmegreen et al.~1993) condensations is
observationally not as pronounced as in the models. The creation of
SSCs or GCs (case~2) in tidal tails is not understood, as it is
generally thought that high gas densities are necessary to produce such
dense clusters. It is not clear, how the intermediate case (3) relates
to the other phenomena. Is is possible that some of these knots merge
to form a massive TDG, or are they dispersed quickly by the tidal
field?

\section{Star formation and morphology of TDGs}
\begin{figure}
  \centerline{\vbox{
      \psfig{figure=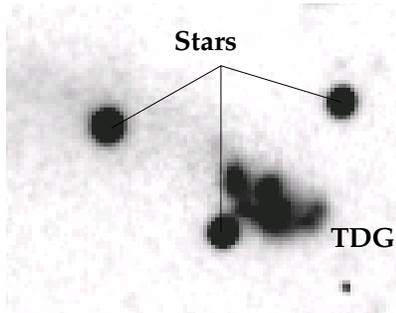,width=0.4\linewidth,angle=0}
  }}
  \caption{$B$-band image of the TDG in AM 1159-530.}
\end{figure}

TDGs do actively form stars and contain molecular gas, but what are
the morphological properties of this star formation?  From optical
observations, especially the highly resolved HST images (Knierman et
al.~2001), one could classify two different types of TDGs, from the
viewpoint of star formation and HII regions:
\begin{enumerate}
\item[a.] {\bf Dense and knotty}: Several clusters or HII regions are
  visible in TDGs. Examples: AM 1159-530 (Fig.~2), NGC 7252W.
\item[b.] {\bf Large and LSB}: A few clusters are seen on an extended
  very low surface brightness object, where the association of the
  clusters to the objects may also be accidental. Example: NGC 7252E.
\end{enumerate}
Type a shows internal kinematics and therefore represents true TDGs,
while this has not been shown without ambiguity for type~b, where more
and deeper spectroscopic data is needed to prove decoupled dynamics.
Although the morphology of these two types is quite different, their
colors are similarly blue, and their spectra both show strong emission
lines, implying comparable star formation activity.  A possible
evolutionary connection between these two ``classes'' of TDGs must
remain speculative for the moment.

\section{Conclusions}
We see 3 modes of young structure in tidal tails, and two types of
TDGs. It is currently not known, what effects are responsible for
these different structures. Especially the possibility of forming
GCs/SSCs in tidal tails is uncertain. New dynamical models are
therefore needed, which provide high enough resolution to resolve
small scale structures along the tails and in the TDGs themselves, and
possibly include SF and feedback to study how SF takes place in TDGs,
and which repercussions stellar winds and supernovae have on the
(future) evolution of TDGs.

From the observational point of view spectrophotometric analysis of
TDG (candidates) is needed, to confirm more `knots' as TDGs. Further
one needs to constrain total and stellar mass, and the burst
properties and possible fading using a wider wavelength range.
Integral field spectroscopy is needed to measure internal
kinematics of TDGs and assess their stability.\\

{\bf Acknowledgements}. PMW thanks the Astronomische Gesellschaft and
the organizers for generous support. We also thank J.E.~Hibbard for
stimulating discussion.

\section*{Discussion}

\noindent{\it B.~Elmegreen:\, } There are two different processes by 
which tidal tails can make dwarf galaxies, on in which the tidal arm
is unstable to collapse along its length, making beads of clusters as
in a spiral arm, and another is the formation of a giant HI cloud at
the end of the arm as a result of movement of the entire outer HI
disk. In addition all the clusters initially at intermediate radius in
the pre-collision galaxy, will end up in the tidal arm too, because of
the tidal motions. \\
\noindent{\it P.~Weilbacher:\, } I agree that in principle one can 
explain how these different structure can be formed in tidal tails.
But simulations only show one of these effects at a time, while in
real interacting systems, like e.g.~NGC 7252, we see two or more
different ``types'' of TDGs. Models can currently not explain why we
see which structures in which tails created during which kind of
encounter geometry. That's why I think we need dynamical
high resolution models. \\

\noindent{\it S.~Sakhibov:\, } What is the range of metallicities in 
studied objects? \\
\noindent{\it P.~Weilbacher:\, } TDGs have metallicities of 
$Z \approx 1/3 Z_\odot$ or $\log O/H \approx 8.5$.

\noindent{\it T.~Armandroff:\, } The galaxies of the Local Group and 
nearby groups define a relation between absolute magnitude and mean
metallicity, with very few outliers. Are you using this relation as a
diagnostic for tidal dwarfs? And can you use this relation to
establish how common surviving tidal dwarfs are? \\
\noindent{\it P.~Weilbacher:\, } Yes, the known (candidates of) TDGs 
have a more or less constant metallicity of the value cited before,
and do not follow the luminosity-metallicity relation. This is used to
discriminate TDGs from possible pre-existing dwarfs in interacting
systems, and should in principle allow to see, if a dwarf galaxy has
tidal origin. \\

\noindent{\it E.~Telles:\, } What is the fate of these condensations 
formed in these tails predicted by the models? Will they fall back
into the system?  Also a comment. It is unlikely that this is a major
mechanism for dwarf galaxy formation because dwarf galaxies are metal
poor and TDGs are not. And we (Telles \& Terlevich 1995) find no
evidence for giant companions around star forming dwarfs such as HII
galaxies. \\
\noindent{\it P.~Weilbacher:\, } In the general models of Barnes \& Hernquist (1992)
and Elmegreen et al.~1993 the models stop shortly after having
developed the condensations. But in the modelling of NGC 7252 (Hibbard
\& Mihos 1995) it is noted that the outer half of the tails can escape
or reach orbits of several Gyrs around the central merger. \\
About the comment: We do not expect that the major number of dwarf
galaxies is formed as TDGs. But your sample and other samples are
preselected to include only galaxies that show the faint emission line
[OIII]4363. It is therefore preselected for metal poor galaxies, and
the missing TDGs might be a selection effect. We also expect TDGs not
to keep up their current SFR for a long time, but instead they could
fade by several magnitudes within one Gyr and then they will not be
detected in surveys, especially emission line surveys will miss them.

\end{document}